%
%
\long\def\UN#1{$\underline{{\vphantom{\hbox{#1}}}\smash{\hbox{#1}}}$}
\def\NP{\vfil\eject}
\def\NI{\noindent}

\magnification=\magstep 2
\overfullrule=0pt
\hfuzz=16pt
\voffset=0.0 true in
\vsize=8.8 true in
\baselineskip 20pt
\parskip 6pt
\hoffset=0.1 true in
\hsize=6.3 true in
\nopagenumbers
\pageno=1
\footline={\hfil -- {\folio} -- \hfil}
 
\noindent{\tt Journal of Statistical Physics 91 (1998) 787-799}

\ 

\ 

\centerline{\UN{\bf ADIABATIC\hphantom{g.}DECOHERENCE}}
 
\ 

\centerline{{\bf Dima Mozyrsky \ {\rm and} \ Vladimir Privman}}

\ 
 
\centerline{\sl Department of Physics, Clarkson University}
\centerline{\sl Potsdam, NY 13699--5820, USA}

\ 
 
\ 

\ 

\NI {\bf KEY WORDS:}\  Quantum decoherence, heat bath, effects of
environment, dissipation
 
\NP

\centerline {\bf ABSTRACT} 

\ 

We study a general quantum system interacting with environment modeled
by the bosonic heat bath of Caldeira and Leggett type. General
interaction Hamiltonians are considered that commute with the system's
Hamiltonian so that there is no energy exchange between the system and
bath. We argue that this model provides an
appropriate description of adiabatic quantum
decoherence, i.e., loss of entanglement on time scales short compared 
to those of thermal relaxation processes associated with energy exchange
with the bath.
The interaction Hamiltonian is then proportional to
a conserved ``pointer
observable.''\  Calculation of the
elements of the reduced density matrix of the system is
carried out exactly, and time-dependence of
decoherence is identified,
similar to recent results for related models. Our key finding
is that the decoherence process is controlled by spectral properties of
the interaction rather than system's Hamiltonian.
  
\NP

\NI{\bf 1.\ \ Introduction}
 
\ 

Quantum decoherence, dissipation, and thermalization due to
interactions with environment have long been important fundamental
issues theoretically and experimentally.$^{(1-11)}$ \ Decoherence
and related topics
have attracted much interest recently due to rapid development of new
fields such as quantum computing and quantum information
theory.$^{(12-18)}$ \ Decoherence due to external interactions is a
major obstacle in the way of implementation of devices such
as quantum computers. Thus in addition to studies of the physics of
decoherence processes there emerged a new field of quantum error
correction$^{(19-25)}$ aiming at effective stabilization of quantum
states against decoherence essentially by involving many additional
quantum systems and utilizing redundancy. The present work contributes
to the former topic: the physics of decoherence.

Decoherence is a result of the coupling of the quantum system under
consideration to the environment which, generally, is the rest of the
universe. In various experimentally relevant situations the interaction
of the quantum system with environment is dominated by the system's
microscopic surroundings. For example, the dominant source of such
interaction for an atom in an electromagnetic cavity is the
electromagnetic field itself coupled to the dipole moment of the
atom.$^{(26)}$ \ In case of Josephson junction in a magnetic
flux$^{(27)}$ or defect propagation in solids, the interaction can be
dominated by acoustic phonons or delocalized electrons.$^{(28)}$ \ 
Magnetic macromolecules interact with the surrounding spin environment
such as nuclear spins.$^{(18)}$ \ Numerous other specific examples could
be cited.

In this work we aim at a general phenomenological description that
models the physically important effects of external interactions as
far as adiabatic decoherence, to be defined later, is concerned. We
note that generally thermalization
and decoherence are associated with the interaction of the quantum
system, described in isolation by the Hamiltonian $H_S$, with another,
large system which we will term the ``bath'' and which internally has
the Hamiltonian $H_B$. The actual interaction will be represented by the
Hamiltonian $H_I$ so that the total Hamiltonian of the system, $H$, is

$$ H=H_S + H_B + H_I \, . \eqno(1.1) $$

\NI It is important to realize that typically the bath is a large,
macroscopic system. Truly irreversible interactions of a quantum system
with its environment, such as thermal equilibration or decoherence
associated with measurement processes, can only be obtained in the
Hamiltonian description (1.1) when it is supplemented by taking 
the limit
of the number of particles or degrees or freedom of the bath going to
infinity.

Interactions of a quantum system with macroscopic systems can
lead to different outcomes. For instance, interaction with a true
``heat bath'' leads to thermalization: the reduced density matrix of
the system approaches $\exp\left(-\beta H_S\right)$ for large times.
Here 

$$ \beta = 1/(kT) \, \eqno(1.2) $$

\NI as usual, and by ``reduced'' we mean the density matrix traced over
the states of the bath. On the other hand for decoherence we expect the
reduced density matrix to approach a diagonal form in the ``preferred
basis'' somehow selected by the ``pointer observable'' Hermitian
operator$^{(1-6,29,30)}$ which is thereby ``measured'' by the macroscopic
system (bath). 

It is important to realize that study of decoherence in the present context
does not fully resolve the problem of understanding quantum measurement
and other fundamental issues at the borderline of quantum and classical
behaviors, such as, for instance, the absence of macroscopic
manifestations of Schr\"odinger-cat type quantum superposition
of states. The more optimistic recent literature$^{(4-6)}$
considers description of 
entanglement and decoherence the key to such
understanding. However,
these fundamental problems have remained open thus far.  

The most explored and probably most tractable approach to modeling
the environmental interactions has
involved representing their effects by coupling the original quantum
system to a set of noninteracting harmonic oscillators (bosonic heat
bath).$^{(1,2,8-11,14,31-33)}$ \ Fermionic heat bath can be also considered,
e.g., Ref.\ 34. We will use the term ``heat bath'' for such systems 
even when they are
used for other than thermalization studies
because they have the temperature parameter defined via initial
conditions, as described later. 

Rigorous formulation of the bosonic heat bath approach was initiated by
Ford, Kac and Mazur$^{(32)}$ and more recently by Caldeira and
Leggett.$^{(11,29)}$ \ It has been established, for harmonic
quantum systems,
that the influence of the heat bath described by the oscillators is
effectively identical to the external uncorrelated random force acting
on a quantum system under consideration. In order for the system to
satisfy equation of motion with a linear dissipation term in the
classical limit the coupling was chosen to be linear in coordinates
while the coupling constants entered lumped in a spectral function
which was assumed to be of a power-law form in the oscillator
frequency, with the appropriate Debye cutoff. We will make this concept
more explicit later.

This model of a heat bath was applied to studying effects of
dissipation on the probability of quantum tunneling from a metastable
state.$^{(8,29)}$ \ It was found that coupling a quantum 
system to the heat bath actually decreases the quantum tunneling rate.
The problem of a particle in a double well potential was also
considered.$^{(9,33)}$   
In this case the interaction with the bath leads to 
quantum coherence loss and complete localization at zero temperature.
This study has lead to the spin-boson Hamiltonian$^{(9,10)}$ which
found numerous other applications. The Hilbert
space of the quantum systems studied was effectively restricted to the
two-dimensional space corresponding to the two lowest energy levels.

Another possible application of the bosonic
heat bath
model concerns aspects of quantum measurement. It is believed that the 
bath is an intrinsic part of a measuring device. In other words, it 
continuously monitors the physical quantity whose operator is coupled 
to it.$^{(4-6)}$ It has been shown in the exactly solvable
model of the quantum oscillator coupled to a heat bath$^{(5)}$ 
that the reduced density matrix of the quantum system 
decoheres, i.e., looses its off-diagonal elements representing the
quantum correlations in the system, in the eigenbasis of the interaction
Hamiltonian. It has also been argued that the
time scale on which this ``measurement'' occurs is much less than
the characteristic time for thermal relaxation of the system.

It is natural to assume that if such a ``bath'' description of the process
of measurement of a Hermitian operator $\Lambda_S$ exists, then the
interaction Hamiltonian $H_I$ in (1.1) will involve  $\Lambda_S$ as
well as some bath-Hilbert-space operators. No general description of
this process exists. Furthermore, when we are limited to specific
models in order to obtain tractable, e.g., analytically solvable,
examples, then there is no general way to separate decoherence and
thermalization effects.
We note that thermalization is naturally associated with exchange of
energy between the quantum system and heat bath.
Model system results and general expectations mentioned
earlier suggest that at least in some cases decoherence
involves its own time scales which are shorter than those of approach
to thermal equilibrium.

In this work we propose to study \UN{adiabatic
decoherence}, i.e., a special case of no energy exchange between the
system and bath. Thus we assume that $H_S$ is conserved, i.e., $[H_S,H]=0$.
This assumption is a special case of ``quantum nondemolition measurement''
concept$^{(2,30)}$ exemplified by the Kerr effect, for instance.
Since $H_S$ and $H_B$ is (1.1) operate in different Hilbert spaces,
this is equivalent to requiring

$$ [H_S,H_I]=0 \, . \eqno(1.3) $$

\NI Furthermore, we will assume that $H_I$ is linear in $\Lambda_S$:

$$ H_I=\Lambda_SP_B \, , \eqno(1.4) $$

\NI where $P_B$ acts in the Hilbert space of the bath. Then
we have 

$$ [\Lambda_S,H_S]=0 \, . \eqno(1.5) $$

\NI Thus, we consider cases in which the measured,
``pointer'' observable 
$\Lambda_S$ is one of the conserved quantities of the quantum system
when it is isolated. Interaction with the
bath will then correspond to
measurement of such an observable, which can be the energy itself.
Specifically, the model of Ref.~14
corresponds to $\Lambda_S = H_S$ for the case of the spin-$1\over
2$ two-state system, motivated by quantum-computing applications;
see also Refs.\ 2, 12-15. The models of Refs.~1 and 2
correspond to the choices of $\Lambda_S = H_S$ and
$\Lambda_S = f(H_S)$, respectively, for a system
coupled to a bosonic spin bath, where $f$ is an arbitrary well-behaved function.

Here we derive exact results for adiabatic decoherence due to coupling
to the bosonic heat bath, assuming general $\Lambda_S$
that commutes with $H_S$. While technically this
represents an extension of the results of Refs.~1 and 2,
we demonstrate that the general case reveals certain
new aspects of the decoherence process. Our new
exact-solution method utilizes coherent states
and may be of interest in other applications as well. In Section 2,
we define the system. Specifically, we choose the bosonic heat bath
form for $H_B$ and $P_B$ in (1.1) and (1.4), but we keep
$H_S$ and $\Lambda_S$ general. However,
we also analyze the mechanism leading to exact
solvability of general models of this type.
Section 3 reports our derivation of the 
exact expression for the reduced density matrix of the system.
Discussion of the results and definition of the continuum limit are
given in Section 4.
 
\NP
 
\NI{\bf 2.\ \ Models of Adiabatic Decoherence}
 
\ 
 
We will be mainly interested in 
the following Hamiltonian for the quantum system
coupled to a bath of bosons (harmonic oscillators)
labeled by the subscript $k$:

$$H=H_S+\sum_k \omega_ka_k^{\dag}a_k+
\Lambda_S \sum_k\left( g_k^*a_k+g_k a_k^{\dag}\right)
\, . \eqno(2.1)$$

\NI Here $a^{\dag}_k$ and $a_k$ are bosonic creation and annihilation 
operators, respectively, so that their commutation relation is
$[a_k,a^{\dag}_k]=1$. The second 
term in (2.1) represents the free field 
or Hamiltonian of the heat bath $H_B$. The last term is the
interaction Hamiltonian $H_I$. The coupling constants will be 
specified later; exact results obtained in Section~3 apply for
general $\omega_k$ and $g_k$. Here and in the following we 
use the convention 

$$\hbar=1 \eqno(2.2)$$

\NI and we also assume that the energy 
levels of each oscillator are shifted by ${1 \over 2}\omega_k$ so
that the ground state of each oscillator has zero energy.

Since we assume that $H_S$ and
$\Lambda_S$ commute, we can select a common
set $|i\rangle$ of eigenstates:

$$ H_S|i\rangle = E_i|i\rangle \, , \eqno(2.3)$$

$$ \Lambda_S |i\rangle = \lambda_i |i\rangle \, . \eqno(2.4) $$

\NI One of the simplifications here, due to the fact that $H_S$ and
$\Lambda_S$ commute, is that these eigenstates
automatically constitute the ``preferred basis'' mentioned
earlier. 

We will assume that initially the quantum system is in a pure 
or mixed state described by the density matrix $\rho (0)$, 
not entangled with the bath. For the bath, we assume that
each oscillator is independently thermalized (possibly by prior
contact with a ``true'' heat bath) at temperature $T$, with the
density matrix $\theta_k$. The total system-plus-bath density
matrix will then be the product

$$ \rho (0) \prod_k \theta_k \, . \eqno(2.5) $$

\NI Here 

$$ \theta_k= Z_k^{-1} e^{-\beta
\omega_k a_k^{\dag}a_k} \, , \eqno(2.6) $$

$$ Z_k\equiv (1-e^{-\beta \omega_k})^{-1} \, , \eqno(2.7) $$

\NI where $Z_k$ is the partition function for the oscillator $k$.
The quantity $\beta$ was defined in (1.2). Introduction of the
temperature parameter via the initial state of the bath is common
in the literature.$^{(1,2,8-11,14-17,29,32,33)}$ \ While it may seem artificial,
we recall that the bath is supposed to be a large system presumably
remaining thermalized on the time scales of interest. Specific results
indicating that the bosonic heat bath can be viewed as a source
of thermalizing noise have been mentioned earlier; see also Ref.\ 35.

Our objective is to study the reduced density matrix of the system
at time $t\geq 0$; it has the following matrix elements in the
preferred basis:

$$ \rho_{mn}(t)={\rm Tr}_B\left[\langle
m|e^{-i {H}t} \left( {\rho}(0) \prod_k \theta_k \right) e^{i {H}t}|n
\rangle\right] \, . \eqno(2.8) $$

\NI Here the outer trace is taken over 
the states of the heat bath, i.e.,
the bosonic modes. The inner matrix element
is in the space of the quantum system.
Note that for no coupling to the
bath, i.e., for $g_k=0$, the density matrix of
the system is simply

$$ \left[\rho_{mn}(t)\right]_{g_k=0}=
\rho_{mn}(0)e^{i(E_n-E_m)t} \, . \eqno(2.9) $$

For the interacting system, the heat-bath states
must be summed over in the trace
in (2.8). It is instructive to consider a more general case
with the bath consisting of independent ``modes'' with the Hamiltonians
$M_k$, so that

$$ H_B=\sum_k M_k \, , \eqno(2.10) $$

\NI where for the bosonic bath we have $M_k=\omega_k a_k^{\dag}a_k$.
Similarly, for the interaction term we assume coupling to each mode independently,

$$ H_I=\Lambda_S \sum_k J_k \, , \eqno(2.11) $$

\NI where for the bosonic bath we have $J_k=g_k^*a_k+ga_k^{\dag}$.
Relation (2.5) remains unchanged, with the definitions
(2.6) and (2.7) replaced by

$$ \theta_k=Z_k^{-1} e^{-\beta M_k}  \, , \eqno(2.12) $$

$$ Z_k = {\rm Tr}_k \left[ e^{-\beta M_k} \right]  \, , \eqno(2.13) $$

\NI where the trace is over a single mode $k$.

Owing to the fact that $H_S$ and $\Lambda_S$ share
common eigenfunctions,
the inner matrix element
calculation in (2.8), in the
system space, can be expressed in terms of the
eigenvalues defined in (2.3)-(2.4). Specifically, we define the bath-space
operators

$$ h_i=E_i + \sum_k M_k
+\lambda_i\sum_k J_k \, , \eqno(2.14) $$

\NI which follow from the form of the Hamiltonian. The calculation
in (2.8) then reduces to

$$ \rho_{mn}(t)={\rm Tr}_B \left[ \langle m|e^{-ih_mt}\left( 
\rho(0)\prod_k \theta_k \right) e^{ih_nt}
|n\rangle \right] \, , \eqno (2.15)$$

\NI which yields  the expression

$$\rho_{mn}(t)=\rho_{mn}(0){\rm Tr}_B
\left[e^{-ih_mt}\left(\prod_k \theta_k
\right)e^{ih_nt}\right] \, . \eqno (2.16) $$

We will now assume that the operators of different modes $k$
commute. This is obvious for the bosonic or spin baths
and must be checked explicitly if one
uses the present formulation for a fermionic
bath. Then we can factor the expression (2.16) as follows:

$$\rho_{mn}(t)=\rho_{mn}(0)e^{i(E_n-E_m)t} \prod_k \left\{
{\rm Tr}_k \left[ e^{-i(M_k+\lambda_m J_k)t}
\theta_k e^{i(M_k+\lambda_n J_k)t} \right] \right\} \, . \eqno (2.17)$$

\NI This expression, or variants derived in earlier works,$^{1,2,14}$
suggests that the problem is exactly solvable
in some cases. Indeed, the inner trace
in over a \UN{single mode} of the bath. For a spin bath of
spin-$1\over 2$ ``modes'' the calculation involves only
$(2 \times 2)$-matrix manipulations and is therefore
straightforward.$^{2,14}$\  However, in this case the
only nontrivial choice of the ``pointer observable''
corresponds, in our notation, to $\Lambda_S=H_S$,
with both operators usually chosen equal to the Pauli
matrix $\sigma_z$. There is also hope for obtaining
analytical results for other baths with modes in finite-dimensional
spaces, such as spins other than $1\over 2$; we have not explored this possibility.

For the bosonic spin bath, the calculation is in the space of a
single harmonic oscillator. It can be carried out by using operator
identities.$^{1,2}$\  We have used instead a method based on the
coherent-state formalism which is detailed in the next section. 

\NP

\NI{\bf 3.\ \ Exact Solution for the Density Matrix}

\ 

We utilize the coherent-state formalism, e.g., Refs.\ 35, 36. 
The coherent states $|z\rangle$ are
the eigenstates of the annihilation
operator $a$ with complex eigenvalues $z$. 
Note that from now on we omit the oscillator
index $k$ whenever this leads to no confusion. 
These states are not orthogonal:

$$ \langle z_1|z_2\rangle =\exp{\left( z_1^*z_2-{1\over 2}|z_1|^2-
{1\over 2}|z_2|^2\right)}  \, . \eqno(3.1) $$   

\NI They form an over-complete set, and one can
show that the identity operator
in a single-oscillator space can be obtained as the integral

$$\int d^2z \, |z\rangle\langle z|=1 \, . \eqno(3.2) $$

\NI Here the integration by definition corresponds to

$$ d^2z \equiv {1\over \pi}d\left({\rm Re}z\right) 
d\left({\rm Im}z\right)  \, . \eqno(3.3) $$

\NI Furthermore, for an arbitrary operator $A$, we have,
in a single-oscillator space, 

$$ {\rm Tr} A = \int d^2z \, \langle z|A|z\rangle  \, . \eqno(3.4) $$

\NI Finally, we note the
following identity,$^{(35)}$ which will be used later,

$$e^{\Omega a^{\dag}a}={\cal
N}\left[e^{a^{\dag}(e^{\Omega}-1)a}\right]   \, . \eqno(3.5) $$

\NI In this relation $\Omega$ is
an arbitrary c-number, while ${\cal N}$ 
denotes normal ordering.

The result (2.17) for the reduced density matrix,
assuming the bosonic spin bath, can be written as

$$\rho_{mn}(t)=\rho_{mn}(0)e^{i(E_n-E_m)t} \prod_k
{S_{mn,k}} \equiv \left[\rho_{mn}(t)\right]_{g_k=0}  \prod_k
{S_{mn,k}} \, , \eqno (3.6)$$

\NI where we used (2.9). Omitting the mode index $k$ for simplicity,
the expression for $S_{mn}$ for each mode in the product is

$$ S_{mn}=Z^{-1}{\rm Tr}\left[e^{-it\gamma_m}e^{-\beta\omega a^{\dag}a}
e^{it\gamma_n}\right] \, , \eqno (3.7)$$

\NI where the trace is in the space of that mode, and we defined

$$\gamma_m=\omega a^{\dag}a +\lambda_m\left(g^* a +
g a^{\dag}\right) \, . \eqno (3.8)$$

\NI The partition function $Z$ is given in (2.7). Relations (3.6)-(3.8)
already illustrate one of our main results: apart from the phase factor which
would be present in the noninteracting case anyway, the system energy
eigenvalues $E_n$ do not enter in the expression for $\rho_{mn}(t)$.
The interesting time dependence is controlled by the eigenvalues
$\lambda_n$ of the ``pointer observable'' operator $\Lambda_S$ (and by
the heat-bath coupling parameters $\omega_k$ and $g_k$).

In order to evaluate the trace in (3.7), we use the coherent-state
approach. We have

$$Z S_{mn}=\int d^2z_0\, d^2z_1\ d^2z_2 \ \langle
z_0|e^{-it\gamma_m}|z_1\rangle
\langle z_1|e^{-\beta\omega a^{\dag} a}|z_2\rangle \langle z_2|
e^{it\gamma_n}|z_0\rangle \, . \eqno (3.9)$$

\NI The normal-ordering formula (3.5) then yields for the middle term,

$$\langle z_1|e^{-\beta\omega a^{\dag} a}|z_2\rangle = 
\langle z_1|z_2\rangle e^{z_1^*(e^{-\beta\omega}-1)z_2}=$$

$$\exp{\left[z_1^*z_2-{1\over
2}|z_1|^2-{1\over 2}|z_2|^2 +z_1^*(e^{-\beta 
\omega} -1)z_2\right]} \, . \eqno (3.10)$$

In order to evaluate the first and last factors in (3.9) we define
shifted operators

$$ \eta = a + \lambda_m { \omega^{-1}g} \, , \eqno (3.11)$$

\NI in terms of which we have

$$\gamma_m=\omega\eta^{\dag}\eta- \lambda_m^2 \omega^{-1} |g|^2
\, . \eqno (3.12)$$

\NI Since $\eta$ and $\eta^{\dag}$ still satisfy the bosonic
commutation relation $[\eta,\eta^{\dag}]=1$, the normal-ordering
formula applies. Thus, for the first factor in (3.9), for instance,
we get

$$\langle z_0|e^{-it\gamma_m}|z_1\rangle=e^{it\lambda_m^2
{|g|^2\over \omega}}\langle z_0|z_1\rangle e^{
\left(e^{-i\omega t}-1\right) \left(z_0^*+\lambda_m
{g^*\over \omega}\right)\left(z_1+\lambda_m{g\over \omega}\right)}
 \, . \eqno (3.13)$$

Collecting all these expressions, one concludes that
the calculation of $S_{mn}$ involves six Gaussian integrations
over the real and imaginary parts of the variables $z_0, z_1,
z_2$. This is a rather lengthy calculation but it can be carried out
in closed form. The result, with indices $k$ restored, is

$$ S_{mn,k} = \exp{ \left( - \omega_k^{-2} |g_k|^2 P_{mn,k} \right) }
\, , \eqno(3.14) $$

\NI where

$$P_{mn,k}=2\left(\lambda_m-\lambda_n\right)^2
\sin^2 {\omega_k t\over 2}
\coth{{\beta\omega_k \over 2}}+i
\left(\lambda_m^2-\lambda_n^2\right)\left(\sin{\omega_kt}-\omega_k
t\right) \, . \eqno (3.15)$$

The expression (3.15), with (3.14), when inserted in (3.6),
is the principal result of this
section. It will be discussed in the next section. Here we note
that in the studies of systems involving the bosonic
heat bath one frequently adds the
``renormalization'' term$^{2,29}$ in the Hamiltonian,

$$ H=H_S+H_B+H_I+H_R \, , \eqno(3.16) $$

\NI where in our case

$$ H_R=\Lambda_S^2\sum_k \omega_k^{-1} |g_k|^2 \, . \eqno(3.17) $$

\NI The role of this renormalization has been reviewed in
Ref.~29. Here we only notice that the sole effect
of adding this term in our calculation is to modify
the imaginary part of $P_{mn,k}$ which plays no role
in our subsequent discussion. The modified expression is

$$P_{mn,k}=2\left(\lambda_m-\lambda_n\right)^2
\sin^2 {\omega_k t\over 2}
\coth{{\beta\omega_k \over 2}}+i
\left(\lambda_m^2-\lambda_n^2\right) \sin{\omega_kt} \, . \eqno (3.18)$$

\NP 

\NI{\bf 4.\ \ Continuum Limit and Discussion} 

\ 

The results of the preceding section, (3.6), (3.14), (3.15),
can be conveniently discussed if we consider magnitudes of
the matrix elements of the reduced density matrix $\rho (t)$. We have

$$ | \rho_{mn} (t) | = | \rho_{mn} (0) | \exp\left[ -{1\over 4} 
\left(\lambda_m-\lambda_n\right)^2 \Gamma (t) \right] \, , \eqno(4.1) $$

\NI where we introduced the factor ${1\over 4}$ to have the
expression identical to that obtained in Ref.\ 14:

$$ \Gamma (t) = 8 \sum_k \omega_k^{-2} |g_k|^2
\sin^2 {\omega_k t\over 2}
\coth{{\beta\omega_k \over 2}} \, . \eqno(4.2) $$

\NI These results suggest several interesting conclusions. First,
the decoherence is clearly controlled by the interaction with
the heat bath rather than by the system's Hamiltonian. The
eigenvalues of the ``pointer observable'' $\Lambda_S$ determine
the rate of decoherence, while the type
of the bath and coupling
controls the form of the function $ \Gamma (t) $.
It is interesting to note that states with equal eigenvalues
$\lambda_m$ will remain entangled even if their energies $E_m$
are different. As expected, the magnitude of the diagonal matrix
elements remains unchanged.

Secondly, we note that $ \Gamma (t) $ is a sum of
positive terms. However, for true decoherence, i.e., in
order for this sum to diverge for large times,
one needs a \UN{continuum} of frequencies and interactions with
the bath modes that are
strong enough \UN{at low frequencies}; see below. From this point
on, our discussion of the function $\Gamma (t) $
is basically identical to that in Ref.\ 14 (see also Ref.~1); we
only outline the main points.
In the continuum limit, exemplified for instance by phonon
modes in solid state, we  introduce the density of states
$G(\omega )$ and sum over frequencies rather than modes
characterized by their wave vectors.
The latter change of the integration variable introduces the
factor which we will loosely write as $dk \over d\omega $; it
must be calculated from the dispersion relation of the bosonic
modes. Thus we have

$$ \Gamma (t) \propto \int d \omega \, {dk \over d\omega} G(\omega ) 
 |g (\omega )|^2 \, \omega^{-2} \sin^2 {\omega t\over 2}
\coth{{\beta\omega \over 2}} \, . \eqno(4.3) $$

\NI In Ref.\ 14, the following choice was considered, motivated
by properties of the
phonon field in solids; see also Refs.\ 8-11, 12-18, 29:

$${dk \over d\omega}G(\omega)|g(\omega)|^2\propto \omega^n e^{-{\omega
\over \omega_c}} \, . \eqno(4.4) $$

\NI This combination of the coupling constants and frequencies
has been termed the spectral function. Here $\omega_c$ is the
Debye cutoff frequency.

Specifically, the authors of Ref.\ 14 have analyzed the cases $n=1$ and $n=3$.
For $n=1$, three regimes were identified,
defined by the time scale for thermal decoherence, $\beta$,
which is large for low temperatures, see (1.2), and the time scale
for quantum-fluctuation effects, $\omega_c^{-1}$. Recall that we use the
units $\hbar=1$. The present treatment only makes sense 
provided $\omega_c^{-1}\ll \beta$. According to Ref.\ 14, the first, 
``quiet'' regime $t \ll \omega_c^{-1}$
corresponds to no significant decoherence and $\Gamma \propto 
(\omega_c t)^2$. The next,
``quantum'' regime, $ \omega_c^{-1} \ll t \ll \beta$,
corresponds to decoherence driven
by quantum fluctuations and $\Gamma \propto \ln (\omega_c t)$.
Finally, for $t \gg \beta$, in the ``thermal'' regime,
thermal  fluctuations play major role in decoherence and $\Gamma \propto
t/\beta $.

For $n=3$, decoherence is incomplete.$^{(14)}$ \ Indeed, while $n$
must be positive for the integral in (4.3) to converge, only for
$n<2$ we have divergent $\Gamma (t)$ growing according to a power
law for large times (in fact, $\propto t^{2-n}$) 
in the ``thermal'' regime. Thus, strong enough coupling $|g(\omega)|$
to the low-frequency modes
of the heat bath is crucial for full decoherence.

In summary, we derived exact results for the model of decoherence
due to energy-conserving interactions with the bosonic heat bath.
We find that the spectrum of the ``pointer observable'' that enters
the interaction with the bath controls the rate
of decoherence. The precise functional form of the time
dependence is determined both by the choice of 
heat-bath and system-bath coupling. However, for the case studied, it
is universal for all pointer observables and for all the 
matrix elements of the reduced density matrix.
 
The authors would like to thank Professor L.\ S.\ Schulman 
for useful discussions. This work has been supported in part by US Air
Force grants, contract numbers F30602-96-1-0276 and F30602-97-2-0089. 
This financial assistance is gratefully acknowledged.

\NP
 
\centerline{\bf References}{\frenchspacing
 
\
 
\item{1.} N.G. van Kampen, J. Stat. Phys. {\bf 78},
299 (1995).

\item{2.} J. Shao, M.-L. Ge and H. Cheng, Phys. Rev. E {\bf 53},
1243 (1996).

\item{3.} Review: H. Brandt, in {\it SPIE
Conf. AeroSence 97}, Proc. Vol. no. 3076, p. 51 (SPIE Publ., 1997).

\item{4.} W. H. Zurek, Physics Today, October 1991, p. 36.

\item{5.} W. G. Unruh, W. H. Zurek, Phys. Rev. D {\bf 40},
1071 (1989).

\item{6.} W. H. Zurek, S. Habib and J. P. Paz,
Phys. Rev. Lett. {\bf 70}, 1187 (1993).

\item{7.} M. Brune, E. Hagley, J. Dreyer, X. Maitre, A. Maali,
C. Wunderlich, J. M. Raimond and S. Haroche, 
Phys. Rev. Lett. {\bf 77}, 4887 (1996).

\item{8.} A. O. Caldeira and A. J. Leggett,
Phys. Rev. Lett. {\bf 46}, 211 (1981).

\item{9.} S. Chakravarty and A. J. Leggett, Phys. Rev. Lett.
{\bf 52}, 5 (1984).

\item{10.} Review: A. J. Legget, S. Chakravarty, A. T. Dorsey,
M. P. A. Fisher and W. Zwerger, Rev. Mod. Phys. {\bf 59}, 1
(1987) [Erratum {\it ibid.\/} {\bf 67}, 725 (1995)].

\item{11.} A. O. Caldeira and A. J. Leggett, Physica {\bf 121A},
587 (1983).

\item{12.} Review: A. Ekert and R. Jozsa, Rev. Mod. Phys. {\bf 68},
733 (1996).

\item{13.} Review: D. P. DiVincenzo, Science {\bf 270}, 255 (1995).

\item{14.} G. M. Palma, K. A. Suominen and A. K. Ekert,
Proc. Royal Soc. London A {\bf 452}, 567 (1996).

\item{15.} W. G. Unruh, Phys. Rev. A {\bf 51}, 992 (1995).

\item{16.} T. Pellizzari, S. A. Gardiner, J. I. Cirac and P. Zoller,
Phys. Rev. Lett. {\bf 75}, 3788 (1995).

\item{17.} A. Garg, Phys. Rev. Lett. {\bf 77}, 764 (1996).

\item{18.} I. S. Tupitsyn, N. V. Prokof'ev, P. C. E. Stamp,
Int. J. Modern Phys. B {\bf 11}, 2901 (1997).

\item{19.} Review: J. Preskill, {\it Reliable Quantum Computers},
preprint (available at http://xxx.lanl.gov/abs/quant-ph/9705031).

\item{20.} Review: D. P. DiVincenzo, {\it Topics in Quantum Computers},
preprint (available at http://xxx.lanl.gov/abs/cond-mat/9612126).

\item{21.} E. Knill and R. Laflamme, Phys. Rev. A {\bf 55}, 900 (1997).

\item{22.} P. Shor, {\it Fault-tolerant quantum computation},
preprint (available at http://xxx.lanl.gov/abs/quant-ph/9605011).

\item{23.} Review: D. Gottesman, {\it Stabilizer Codes and Quantum
Error Correction}, preprint (available at
http://xxx.lanl.gov/abs/quant-ph/9705052).

\item{24.} D. Aharonov and M. Ben-Or, {\it Fault-Tolerant Quantum
Computation with Constant Error}, preprint (available at
http://xxx.lanl.gov/abs/quant-ph/9611025).

\item{25.} A. Steane, {\it Active Stabilisation, Quantum Computation
and Quantum State Synthesis}, preprint (available at
http://xxx.lanl.gov/abs/quant-ph/9611027).

\item{26.} C. W. Gardiner {\it Handbook of Stochastic Methods
for Physics, Chemistry and the Natural Sciences}
(Springer-Verlag, 1990).

\item{27.} A. J. Leggett, in {\it Percolation, Localization and
Superconductivity}, NATO ASI Series B: Physics, Vol. {\bf 109},
edited by A. M. Goldman and S. A. Wolf (Plenum, New York 1984), p. 1.

\item{28.} J. P. Sethna, Phys. Rev. B {\bf 24}, 698 (1981).

\item{29.} Review: A.O. Caldeira and A.J. Leggett,
Ann. Phys. {\bf 149}, 374 (1983).

\item{30.} L. Mandel and E. Wolf, {\it Optical Coherence 
and Quantum Optics}, section 22.6, p.1100 (Cambridge University Press, 1995).

\item{31.} R. P. Feynman and A. R. Hibbs,
{\it Quantum Mechanics and Path Integrals} 
(McGraw-Hill Book Company, 1965).

\item{32.} G. W. Ford, M. Kac and P. Mazur,
J. Math. Phys. {\bf 6}, 504 (1965).

\item{33.} A. J. Bray and M. A. Moore, Phys. Rev. Lett.
{\bf 49}, 1546 (1982).

\item{34.} L.-D. Chang and S. Chakravarty, Phys. Rev. B
{\bf 31}, 154 (1985).

\item{35.} W. H. Louisell, {\it Quantum Statistical
Properties of Radiation}
(Wiley, New York, 1973).

\item{36.} Review: M. Hillery, R. F. O'Connell, M. O. Scully
and E. P. Wigner, Phys. Rep. {\bf 106}, 121 (1984).

}
 
\bye